\documentstyle[aps,prl,psfig]{revtex}
\begin{document}

\title {Low Hubble Constant from Type Ia Supernovae by van
den Bergh's Method}

\author{David Branch, Adam Fisher, E. Baron \& Peter Nugent} 

\address{Department of Physics and Astronomy, University of Oklahoma, Norman,
OK 73019-0225}
\date{\today}
\maketitle

\begin{abstract}
An interesting way to calibrate the absolute magnitudes of remote
Type Ia supernovae (SNe Ia) that are well out in the Hubble flow$^1$,
and thus determine the value of the Hubble constant, $\bf H_0$, has
been introduced by van den Bergh$^2$.  His approach relies on
calculations$^3$ of the peak absolute magnitudes and broad--band
colors for SN Ia explosion models. It does not require any corrections
for extinction by interstellar dust, and no SNe Ia are excluded on
grounds of peculiarity.  Within the last few years distances have been
determined to the parent galaxies of six SNe Ia by means of Cepheid
variables$^{4-10}$. Cepheid--based distances also have become
available for three other SNe Ia if one is willing to use the distance
to a galaxy in the same group in lieu of the distance to the parent
galaxy itself.  Here we determine the value of $\bf H_0$ in a way that is
analogous to that of van den Bergh, but now using Cepheid--based
distances instead of calculated light curves.  We obtain $\bf H_0 = 55
\pm 5\ \rm \bf km\ s^{-1}\ Mpc^{-1}$.  This value, with $\bf \Lambda=0$ and
$\bf \Omega=1$, corresponds to a cosmic expansion time of $\bf 12 \pm
1$ Gyr, which is consistent with several recent determinations of the
ages of globular clusters.
\end{abstract}

\medskip

van den Bergh noted that the explosion--model light curves obey a
relation between the peak visual absolute magnitude $M_V$ and the
$B-V$ color, with a slope that is nearly the same as that of the
standard extinction law, $A_V/E(B-V)=3.1$.  Consequently one can
define a parameter
  
\[ 
M^*_V = M_V - 3.1(B-V)
\]                                

\noindent which is, to a first approximation, independent of both
extinction and supernova model.  From the models, van den Bergh
derived values of $M^*_V$ ranging from $-19.60 \pm 0.05$ to $-19.75
\pm 0.02$ depending on which weights were assigned to the various
models.  For the 13 real SNe Ia in the CTIO remote sample, which is
well out in the Hubble flow ($3000 \le cz \le 30,000\ \rm km\
s^{-1}$), van den Bergh found

\[M^*_V = -19.59 \pm 0.11 + 5 log (H_0/60\ \rm km\ s^{-1}\ Mpc^{-1})
\]

\noindent and therefore obtained values of $H_0$ ranging from $60 \pm 3$ to $55
\pm 3\ \rm km\ s^{-1}\ Mpc^{-1}$.  These values depend on the models
and the light--curve calculations but they are independent of any
astronomical calibration.

\medskip
\begin{center}
\begin{tabular}{rcrrrrr}
\hline
\multicolumn{7}{c}{TABLE 1 Cepheid--calibrated type Ia supernovae}\\\hline
\multicolumn{1}{c}{SN}&    galaxy&
\multicolumn{1}{c}{B}&\multicolumn{1}{c}{V}& \multicolumn{1}{c}{B-V}&
\multicolumn{1}{c}{$\mu$} & \multicolumn{1}{c}{$M_V$}\\ 
1937C & IC 4182 &  8.71$\pm$0.14 &  8.72$\pm$0.06&
-0.03$\pm$0.13 &  28.36$\pm$0.09 & -19.64$\pm$0.11\\ 
1960F & NGC 4496 & 11.58$\pm$0.05 & 11.49$\pm$0.15&
\phantom{-}0.09$\pm$0.16 &  31.10$\pm$0.13& -19.61$\pm$0.20\\ 
1972E & NGC 5253 &  8.61$\pm$0.21 &  8.61$\pm$0.12&
\phantom{-}0.00$\pm$0.09 &  28.08$\pm$0.10& -19.47$\pm$0.16\\ 
1981B & NGC 4536 & 12.04$\pm$0.04 & 11.98$\pm$0.04&
\phantom{-}0.04$\pm$0.06 &  31.10$\pm$0.13& -19.12$\pm$0.14\\ 
1986G & NGC 5128 & 12.45$\pm$0.05 & 11.40$\pm$0.05&
\phantom{-}1.05$\pm$0.07 &  28.08$\pm$0.41& -16.68$\pm$0.42\\ 
1989B & NGC 3627 & 12.34$\pm$0.05 & 11.99$\pm$0.05&
\phantom{-}0.35$\pm$0.07 &  30.32$\pm$0.16& -18.33$\pm$0.17\\ 
1990N & NGC 4639 & 12.70$\pm$0.05 & 12.61$\pm$0.05&
\phantom{-}0.09$\pm$0.07 &  32.00$\pm$0.23& -19.39$\pm$0.24\\ 
1991T & NGC 4527 & 11.64$\pm$0.05 & 11.50$\pm$0.04&
\phantom{-}0.14$\pm$0.06 &  31.10$\pm$0.16& -19.60$\pm$0.16\\\hline 
\end{tabular}
\end{center}

\medskip

Data for the Cepheid--calibrated SNe Ia for which the peak $B$ and $V$
magnitudes are known are listed in Table 1. (SN 1895B in NGC 5253
cannot be used here because only $B$ is known.)  Sources of the data
are as follows.  SN 1937C: $B$ and $V$ are from Schaefer$^{11}$ and
the distance modulus, $\mu$, is from Saha et al.$^6$.  The
uncertainty in $B-V$ is less than would be obtained from the
quadrature sum of the uncertainties in $B$ and $V$ because the
uncertainties in $B$ and $V$ are correlated.  SN 1960F: $B$, $V$, and
$\mu$ are from Saha et al.$^{10}$.  Somewhat different values, $B=11.77
\pm 0.07$ and $V=11.51 \pm 0.18$, have been reported$^{12}$ but
these are based on less information than those of Saha et al. and they
make SN 1960F suspiciously red, with $B-V=0.26$ (although with an
uncertainty of $\pm0.19$).  SN 1972E: $B$ and $V$ are from Hamuy et
al.$^{1}$, the uncertainty in $B-V$ takes into account
that the uncertainties in $B$ and $V$ are correlated, and $\mu$ is
from Saha et al$^7$.  SN 1981B: $B$ and $V$ are from Schaefer$^{13}$ and
$\mu$ is from Saha et al$^9$.  SN 1986G: $B$ and $V$ are from
Phillips et al.$^{14}$ and $\mu$ is equated to that of SN 1972E
because their parent galaxies, NGC 5128 and NGC 5253, are both members
of the Centaurus group, but with an additional uncertainty of $\pm
0.4$ for SN 1986G because these two galaxies are separated by 11.8
degrees on the sky.  (SN 1986G will not enter into our adopted result,
but its very red $B-V$ of 1.05 will help to illustrate the validity of
the procedure.)  SN 1989B: $B$ and $V$ are from Wells et al.$^{15}$
and $\mu$ is equated to that of NGC 3368$^{16}$ because
NGC 3627, the parent galaxy of SN 1989B, and NGC 3368 are fellow
members of the Leo spur$^{17}$.  An additional uncertainty of
$\pm 0.14$ has been included for SN 1989B to allow for possible
differences in distance between NGC 3627 and NGC 3368 and 0.05 has
been added for the HST ``long exposure'' effect$^8$.
SN 1990N: $B$ and $V$ are from Leibundgut et al.$^{18}$ and $\mu$ is
from Sandage et al$^8$.  SN 1991T: $B$ and $V$ are from Phillips
et al.$^{19}$ and $\mu$ is equated to that of SNe 1960F and 1981B
because their parent galaxies are thought to be members of the same
compact group$^{20,21}$).  SN 1991bg is not in Table 1
but it is plotted as a special symbol. $B=14.70 \pm 0.10$ and $V=13.95
\pm 0.02$ are from Leibundgut et al$^{22}$.  SN 1991bg, with its red
$B-V$ of 0.75, is considered only to help illustrate the validity of
the procedure.  It will not be used in the analysis because there is
no Cepheid--based distance to NGC 4374, a Virgo elliptical galaxy.
For the illustration we use $\mu=31.62 \pm 0.35$ obtained$^{21}$ from
the ``SEAM'' spectrum--fitting procedure$^{23-25}$ which gives a
distance to SNe 1981B that is in excellent agreement with its
Cepheid--based distance$^{21}$.

\begin{figure}
\psfig{file=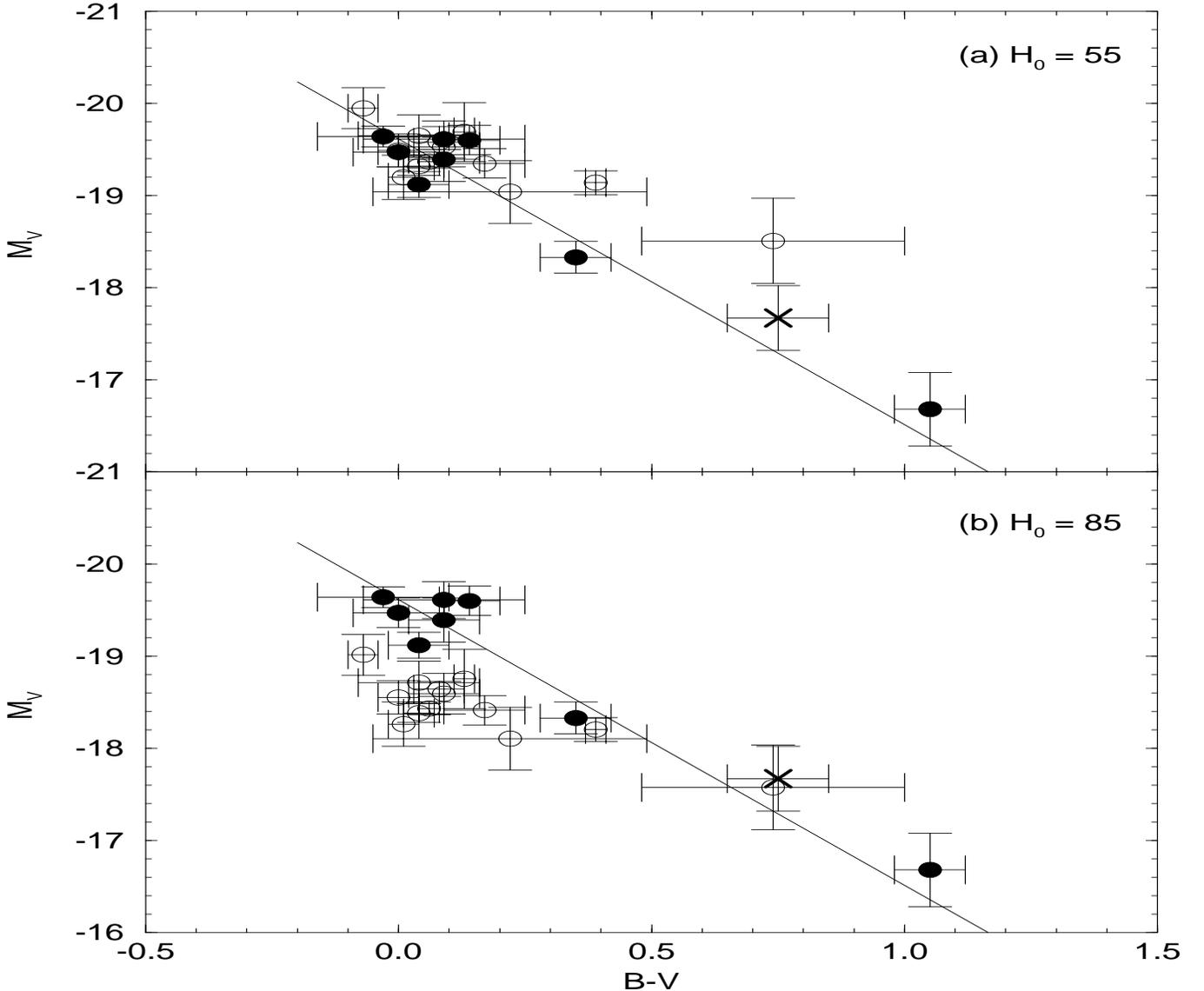,height=0.7\textheight,width=1.0\textwidth}
\caption{{\sl a} Peak visual absolute magnitude, uncorrected
for extinction, is plotted against $B-V$ for Cepheid--calibrated SNe
Ia (filled circles), SN 1991bg (cross), and the CTIO sample of remote
SNe Ia (open circles).  For the latter, $H_0=55\ \rm km\ s^{-1}\
Mpc^{-1}$ has been used.  The straight line has the extinction slope,
$A_V/E(B-V)=3.1$.  {\sl b} Using $\rm H_0=85$ instead of 55.}
\label{fig1}
\end{figure}

$M_V$, uncorrected for extinction, is plotted against $B-V$ in
Fig. 1a.  The eight Cepheid--calibrated SNe Ia are plotted as filled
circles and the 13 SNe Ia of the CTIO remote sample are plotted as open
circles.  The data points tend to lie along the
extinction line.  SN 1989B, intrinsically normal but
extinguished$^{15}$, SN 1991bg, unextinguished but intrinsically dim
and red$^{26,22}$, and SN 1986G, intrinsically dim and red {\sl and}
extinguished$^{14}$, all lie very near the extinction line.  SN 1990Y,
probably intrinsically normal but extinguished$^1$, and SN 1992K,
intrinsically dim and red$^{27}$, are not far off.  The tendency of
real SNe Ia, whether extinguished and/or intrinsically dim and red, to
lie near the extinction line shows that this approach is a useful one.
It also supports the general validity of the explosion--model
light--curve calculations$^3$.

\medskip

Now we turn our attention to the cluster of bright, blue SNe Ia having
$M_V \le -19$ and $B-V \le 0.2$.  These are so tightly clustered that
no correlation between $M_V$ and $B-V$ is readily apparent.  We
suspect that the {\sl intrinsic} properties of these SNe Ia would show
a correlation, because there are good reasons to think that SN 1991T
is significantly extinguished and intrinsically the bluest and
brightest SN Ia yet discovered$^{28}$, and SN 1981B, the faintest of
these clustered data points at $M_V=-19.12$, was mildly extinguished
(e.g., M. M. Phillips, personal communication).  But for the present
procedure, these opinions about the extinction make no difference.  If
we restrict our attention to the clustered data points and match the
11 SNe Ia of the CTIO sample (for which $M^*_V = -19.43 \pm 0.07 + 5
log (H_0/60)$) to the six Cepheid--calibrated SNe Ia (for which $M^*_V
= -19.61 \pm 0.12$), we obtain $H_0=55 \pm 4\ \rm km\ s^{-1}\
Mpc^{-1}$.  Omitting SN 1937C on the grounds that it's $B$ and $V$ are
disputed$^{29,30}$ would not change the result at all, nor would the
omission of the six SNe Ia of the CTIO sample that were discovered
more than 10 days after maximum light. Including the dimmer and redder
SNe 1989B, 1990Y, 1992K, and 1986G, which we do not favor because it
would be stretching the method too far, would give $H_0=58 \pm 3\ \rm
km\ s^{-1}\ Mpc^{-1}$.

\medskip

We adopt $H_0 = 55 \pm 5\ \rm km\ s^{-1}\ Mpc^{-1}$ from this method,
which does not make use of any extinction corrections and does not
entail excluding any SNe Ia on grounds of peculiarity.  Fig. 1b is
just like Fig. 1a except that H$_0$ is set to 85 rather than 55 km
s$^{-1}$ Mpc$^{-1}$.  The few red and dim SNe Ia actually fit better,
but {\sl all} of the bright, blue Cepheid--calibrated SNe Ia are
brighter than {\sl all} of the bright, blue SNe Ia of the CTIO sample.
This is entirely unsatisfactory, and $H_0=85$ is excluded.  To the
extent that there is scatter in the relation between $M_V$ and $B-V$,
and that the CTIO sample of remote SNe Ia is magnitude--selected to a
greater degree than the sample of nearer Cepheid--calibrated SNe Ia,
then the true value of $H_0$ may be a little {\sl lower} than we have
obtained here.

\medskip

van den Bergh$^2$ suggested that the conflict between the low value of
$H_0$ that he obtained using models of SNe Ia, and higher values
obtained by others using Cepheid--based distance determinations to a
few spiral galaxies in the Virgo cluster complex, implied that unless
Cepheids are unreliable distance indicators the models must be either
too blue or too bright.  However, {\sl the present result of $H_0=55
\pm 5\ \rm km\ s^{-1}\ Mpc^{-1}$, obtained by the method of van den
Bergh but using Cepheids rather than models, gives the same low value
of $H_0$.}  This shows that the fault lies not with the models, nor
with the Cepheids, but with the hazardous route through the Virgo
cluster complex$^{21,31}$.

\medskip

The present result of $H_0=55 \pm 5\ \rm km\ s^{-1}\ Mpc^{-1}$ also is
in excellent agreement with detailed spectrum fitting of SNe Ia using
fully relativistic, NLTE calculations$^{24,25,21}$; with methods based
on $^{56}Ni$ radioactivity$^{21}$; with a straightforward
standard--candle treatment$^8$; with a treatment$^{12}$ that takes the
effects of a putative magnitude--decline correlation into account; and
with a result$^{32}$ that distinguishes between SNe Ia in blue and red
galaxies.  The impressive agreement between Cepheid--based and
physically--based calibrations indicates that future modifications to
the Cepheid period--luminosity law will not have strong effects on the
value of $H_0$ obtained from SNe Ia.  It is difficult to see how a
lingering controversy between ``high'' (e.g., 85) and ``low'' (e.g.,
55) values of $H_0$ can be maintained.  The situation is
asymmetric. SNe Ia require a low value and they cannot be reconciled
with a high value.  Which method that is said to favor a high value
cannot be reconciled with the low value that is required by SNe Ia?

\medskip

$H_0=55 \pm 5\ \rm km\ s^{-1}\ Mpc^{-1}$, with $\Lambda=0$ and
$\Omega=1$, corresponds to a universal expansion age of $12 \pm 1$
Gyr, which is consistent with some recent determinations of the ages
of globular clusters$^{33-37}$, and not inconsistent with a recent
estimate of the age of an ultra--metal--poor star based on its content
of radioactive thorium$^{38}$.

\clearpage

\begin{small}

\noindent 1. Hamuy, M., Phillips, M. M., Maza, J., Suntzeff, N. B., Schommer,
R. A., \& Avil\'es, R. {\sl Astr. J.} {\bf 109}, 1 (1995)

\noindent 2. van den Bergh, S. {\sl Astrophys. J.} {\bf 453}, L55 (1995)

\noindent 3. H\"oflich, P. \& Khokhlov, A. {\sl Astrophys. J.} {\bf 457},
500 (1996)

\noindent 4. Sandage, A., Saha, A., Tammann, G. A., Panagia, N., \&
Macchetto, F. D. {\sl Astrophys. J.} {\bf 401}, L7 (1992)

\noindent 5. Sandage, A., Saha, A., Tammann, G. A., Labhardt, L.,
Schwengeler, H., Panagia, N., \& Macchetto, F. D. {\sl Astrophys. J.}
{\bf 423}, L13 (1994)

\noindent 6. Saha, A., Labhardt, L., Schwengeler, H., Macchetto,
F. D., Panagia, N., Sandage, A., \& Tammann, G. A. {\sl Astrophys. J.}
{\bf 425}, 14 (1994)

\noindent 7. Saha, A., Sandage, A., Labhardt, L., Schwengeler, H.,
Tammann, G. A., Panagia, N., \& Macchetto, F. D. {\sl Astrophys. J.}
{\bf 438}, 8 (1995)

\noindent 8. Sandage, A., Saha, A., Tammann, G. A., Labhardt, L.,
Panagia, N., \& Macchetto, F. D. {\sl Astrophys. J.} {\bf 460}, L15 (1996)

\noindent 9. Saha, A., Sandage, A., Labhardt, L., Tammann, G. A.,
Macchetto, F. D.,\& Panagia, N. {\sl Astrophys. J.} (in the press)

\noindent 10. Saha, A., Sandage, A., Labhardt, L., Tammann, G. A.,
Macchetto, F. D.,\& Panagia, N. {\sl Astrophys. J.} (in the press)

\noindent 11. Schaefer, B. E. {\sl Astrophys. J.} {\bf 426}, 493 (1994)

\noindent 12. Schaefer, B. E. {\sl Astrophys. J.} {\bf 460}, L19 (1996)

\noindent 13. Schaefer, B. E. {\sl Astrophys. J.} {\bf 449}, L9 (1995)

\noindent 14. Phillips, M. M., {\sl et al.} {\sl
Publs. Astr. Soc. Pacif.} {\bf 99}, 592 (1987) 

\noindent 15. Wells, L. A., {\sl et al.} {\sl Astr. J.} {\bf 108}, 2233 (1994)

\noindent 16. Tanvir, N. R., Shanks, T., Ferguson, H. C., \& Robinson,
D. T. R. {\sl Nature} {\bf 377}, 27 (1995)

\noindent 17. Tully, R. B. {\sl Astrophys. J.} {\bf 321}, 280 (1987)

\noindent 18. Leibundgut, B., Kirshner, R. P., Filippenko, A. V., Shields,
J. S., Foltz, C. B., Phillips, M. M., \& Sonneborn, G. {\sl
Astrophys. J.} {\bf 371}, L23 (1991)

\noindent 19. Phillips, M. M., Wells, L. A., Suntzeff, N. B., Hamuy, M., 
Leibundgut,B., Kirshner, R. P., \& Foltz, C. B. {\sl Astr. J.} {\bf
103}, 1632 (1992)

\noindent 20. Tully, R. B. {\sl The Catalog of Nearby Galaxies}
(Cambridge University Press 1988)

\noindent 21. Branch, D., Nugent, P., \& Fisher, A. in {\sl Thermonculear
Supernovae} eds. Canal, R., Ruiz--Lapuente, P. \& Isern, J.) (Kluwer,
in the press)

\noindent 22. Leibundgut, B., {\sl et al.} {\sl Astr. J} {\bf 105}, 301 (1993)

\noindent 23. Baron, E., Hauschildt, P. H., \& Branch, D. {\sl
Astrophys. J.} {\bf 426}, 334
(1994)

\noindent 24. Nugent, P., Baron, E., Hauschildt, P. H., \& Branch,
D. {\sl Astrophys. J.} {\bf 441}, L33 (1995)

\noindent 25. Nugent, P., Branch, D., Baron, E., Fisher, A., Vaughan,
T. E., \& Hauschildt, P. H. {\sl Phys. Rev. Lett.} {\bf 75}, 394
(1995); erratum: {\bf 75}, 1874

\noindent 26. Filippenko, A. V., {\sl et al.} {\sl Astr. J.} {\bf
104}, 1543 (1992) 

\noindent 27. Hamuy, M., et al. {\sl Astr. J.} {\bf 108}, 2226 (1994)

\noindent 28. Nugent, P., Phillips, M. M., Baron, E., Branch, D., \&
Hauschildt, P. {\sl Astrophys. J.} {\bf 455}, L147 (1995)

\noindent 29. Pierce, M. J., \& Jacoby, G. H. {\sl Astr. J.} {\bf 110}, 
2885 (1995)

\noindent 30. Schaefer, B. E. {\sl Astr. J.} (in the press)

\noindent 31. Tammann, G. A., Labhardt, L., Federspiel, M., Sandage,
A., Saha, A., Macchetto, F. D., \& Panagia, N. 1996, in {\sl Science
with the Hubble Space Telescope --- II} (eds. Benvenuti, P. Macchetto,
F. D., \& Schreier, E. J) (Space Telescope Science Institute, in the press)

\noindent 32. Branch, D., Romanishin, W., \& Baron, E. {\sl Astrophys. J.}
(in the press)

\noindent 33. Chaboyer, B. {\sl Astrophys. J.} {\bf 444}, L9 (1995)

\noindent 34. Shi, X. {\sl Astrophys. J.} {\bf 446}, 637 (1995)

\noindent 35. Mazzitelli, I., D'Antona, F., \& Caloi, V. {\sl Astr. 
Astrophys.} {\bf 302}, 382 (1995)

\noindent 36. Jimenez, R., Thejll, P., Jorgensen, U., MacDonald, J., \&
Pagel, B. {\sl Mon. Not. R. Astr. Soc.} (in the press)

\noindent 37. Salaris, M., Degl'Innocenti, S., \& Weiss, A. {\sl
Astrophys.  J.} (in the press)

\noindent 38. Sneden, C., McWilliam, A., Preston, G. W., Cowan, J. J.,
Burris, D. L., \& Armosky, J. {\sl Astrophys. J.} (in the press)

\end{small}

\bigskip

\noindent ACKNOWLEDGEMENTS. We thank Allan Sandage, Abi Saha, Gustav
Tammann, and Bradley Schaefer for sending papers in advance of
publication. This work has been supported by NSF.

\end{document}